\documentstyle[epsf]{article}

\newcommand{\beq}{ \begin{equation} }
\newcommand{\eeq}{ \end{equation} }
\newcommand{\bea}{ \begin{eqnarray} }
\newcommand{\eea}{ \end{eqnarray} }
\newcommand{\be}{ \beta }
\newcommand{\f}{ \frac }
\newcommand{\de}{ \partial }
\newcommand{\ex}{ {\rm e} }
\newcommand{\vS}{ {\bf S} }
\newcommand{\ph}{ \varphi }

\begin{document}
\thispagestyle{empty}
\parskip=12pt
\raggedbottom

\def\mytoday#1{{ } \ifcase\month \or
 January\or February\or March\or April\or May\or June\or
 July\or August\or September\or October\or November\or December\fi
%\space\number\day ,
 \space \number\year}
\noindent
\hspace*{9cm} UNICAL-TP 98/2 \\
\vspace*{1cm}
\begin{center}

{\LARGE Fixed point actions and on-shell tree-level Symanzik improvement}

\vspace{1cm}

Alessandro Papa \\
Dipartimento di Fisica, Universit\`a della Calabria \\
Istituto Nazionale di Fisica Nucleare, Gruppo collegato di Cosenza \\
I-87036 Arcavacata di Rende, Cosenza, Italy

\vspace{0.5cm}

%\mytoday \\ \vspace*{0.5cm}

\begin{center}
March 1998
\end{center}
 
\vspace{.5cm}

\nopagebreak[4]

\begin{abstract}
In this paper it is argued that the properties of the fixed point
action of a renormalization group transformation can be used 
to implement the on-shell tree-level Symanzik improvement of lattice 
actions to any given order in the expansion in the lattice spacing, in a  
way which does not involve any perturbative calculations. 
In particular, a well-known technique for the lowest order improvement 
of SU(N) lattice gauge theories is revisited from the point of view
of fixed point actions, which allows to shed light on some subtle points.
\end{abstract}

\end{center}

\vspace{0.5cm}


\vfill

\eject

\section{Introduction}
\label{sec:intro}

The lattice regularization of a field theory provides a unique tool to 
study its non-perturbative properties since it allows to perform
numerical simulations by Monte Carlo techniques. Being the lattice spacing 
finite in numerical simulations, the determination of physical quantities 
from the lattice is plagued by systematic uncertainties. The 
na\"{\i}ve way to circumvent this problem is to adopt in numerical 
simulations the simplest possible lattice action (the so called ``standard 
action'') and to make the lattice spacing smaller and smaller until a safe 
extrapolation to the continuum limit is viable. In practice, this amounts
to consider very fine lattices and, consequently, requires very 
time-consuming and memory-demanding computations, in spite of the 
simplicity of the discretized action. An alternative way consists in 
improving the discretization procedure according to some theoretical 
prescription: the loss of simplicity in the form of the lattice action 
can be largely compensated by the possibility to extract physics from 
numerical simulations on relatively coarse lattices.

The improvement method due to Symanzik~\cite{Sym83} consists in adding
irrelevant terms to the standard action with appropriately 
chosen coefficients to cancel the lattice artifacts in the $n$-point Green 
functions up to a given order in $a^2$ and up to a given order in 
perturbation theory\footnote{In the presence of fermions, lattice artifacts 
appear as power series in $a$, instead of $a^2$. In the following, 
we will restrict for simplicity to bosonic theories.}. For gauge theories 
this program has to be limited to the ``on-shell'' improvement, i.e. to the 
perturbative improvement of the physical quantities~\cite{Wei83,LW85a}.

A more radical improvement can be obtained according to the Wilson 
Renormalization Group (RG) theory~\cite{Wil74}. It is well-established that 
lattice actions lying in the space of the couplings of the theory on the 
fixed point (FP) and on the renormalized trajectory (RT) of a given 
renormalization group transformation are perfect actions~\cite{HN94}, in the 
sense that their spectral properties are completely free of cut-off effects. 
Lattice actions lying on the trajectory which originates at the FP and 
leaves the critical surface in the orthogonal direction are called FP 
actions\footnote{It is implicitly assumed here that there is only one
(weakly) relevant direction, as in the case of 4d SU(N) gauge theories
and of the 2d O(3) $\sigma$-model.}. Their spectral
properties are free of lattice artifacts only in the 
classical limit, i.e. FP actions are {\em classical} perfect actions. 
One could say equivalently that FP actions 
are on-shell tree-level Symanzik improved actions to {\em all} orders in 
$a^2$.

In this paper, it is argued that the properties of FP actions and of their 
classical solutions can be used to understand and to implement the on-shell 
tree-level Symanzik improvement up to any fixed order in $a^2$ in a way
which requires no perturbative 
calculations of on-shell Green functions. In particular, 
a technique for the $O(a^2)$ on-shell tree-level Symanzik 
improvement, which dates back to the arguments given by L\"uscher and 
Weisz at the end of Sect.~5 in Ref.~\cite{LW85b} and has been exploited
for the 4d SU(N) lattice gauge theory in Ref.~\cite{GGSV94} 
and in Ref.~\cite{FP97}, is revisited here from the point of
view of FP actions, which allows to shed light on some subtle points
(see e.g. footnote~(9) of Ref.~\cite{LW85b}). This paper 
develops some statements contained in the Sect.~2 of Ref.~\cite{HN97} where 
the relation between FP actions and on-shell tree-level Symanzik 
improvement was already clearly evidenced.

The paper is organized as follows: in Sect.~2, the $O(a^2)$ on-shell tree
level Symanzik improvement for 4d SU(N) gauge theories according to the 
technique of Refs.~\cite{LW85b,GGSV94} is briefly reviewed; in Sect.~3, the 
relevant properties of perfect actions are recalled;
in Sect.~4, the on-shell tree-level Symanzik 
improvement is considered from the point of view of FP actions,
the lowest order results of Sect.~1 are revisited 
and a recipe is given to implement the improvement to any fixed order 
in $a^2$.

\section{On-shell tree-level Symanzik improvement at $O(a^2)$}
\label{sec:sym}

The result of the tree-level Symanzik improvement is the vanishing of the 
lattice corrections, up to a given order in $a^2$,
in the $n$-point Green functions of a theory. Since the classical action 
coincides with the tree-level generating functional of the proper 
(one-particle-irreducible) Green functions of a theory, one expects that 
lattice
corrections up to that given order should be absent also in the lattice 
action. This suggests that if a lattice action in its general form is 
somehow expanded in a power series of $a^2$, the tree-level Symanzik 
conditions on the coupling constants can be immediately written
by imposing the vanishing of the irrelevant terms 
to a given order in $a^2$. This argument has been indeed exploited 
in 4d SU(N) gauge theories~\cite{GGSV94,FP97} for the on-shell tree-level 
Symanzik improvement at the lowest order in $a^2$.

\begin{figure}[htbp]
\begin{minipage}[t]{54mm}
\begin{center}
\begin{picture}(50,50)(0,0)
\put(0,0){\circle*{4}}
\put(0,0){\line(1,0){40}}
\put(40,0){\circle*{4}}
\put(40,0){\line(0,1){40}}
\put(40,40){\circle*{4}}
\put(40,40){\line(-1,0){40}}
\put(0,40){\circle*{4}}
\put(0,40){\line(0,-1){40}}
\end{picture}\\*[6mm]

plaquette
\end{center}
\end{minipage}
\begin{minipage}[t]{64mm}
\centering
\begin{picture}(90,50)(0,0)
\put(0,0){\circle*{4}}
\put(0,0){\line(1,0){40}}
\put(40,0){\circle*{4}}
\put(40,0){\line(1,0){40}}
\put(80,0){\circle*{4}}
\put(80,0){\line(0,1){40}}
\put(80,40){\circle*{4}}
\put(80,40){\line(-1,0){40}}
\put(40,40){\circle*{4}}
\put(40,40){\line(-1,0){40}}
\put(0,40){\circle*{4}}
\put(0,40){\line(0,-1){40}}
% hidden lines
\multiput(40,0)(0,4){11}{\circle*{0.1}}
%\dashline[30]{3}(40,0)(40,40)
\end{picture}\\*[6mm]

rectangle
\end{minipage}
\begin{minipage}[t]{54mm}
\centering
\begin{picture}(84,76)(0,0)
\put(0,0){\circle*{4}}
\put(0,0){\line(1,0){40}}
\put(40,0){\circle*{4}}
\put(40,0){\line(3,2){26}}
\put(64,16){\circle*{4}}
\put(64,16){\line(0,1){40}}
\put(64,56){\circle*{4}}
\put(64,56){\line(-1,0){40}}
\put(24,56){\circle*{4}}
\put(24,56){\line(-3,-2){26}}
\put(0,40){\circle*{4}}
\put(0,40){\line(0,-1){40}}
% hidden lines
\multiput(0,40)(4,0){11}{\circle*{0.1}}
\multiput(40,40)(3,2){9}{\circle*{0.1}}
\put(40,40){\circle*{4}}
\multiput(40,0)(0,4){11}{\circle*{0.1}}
%\dashline[30]{3}(0,40)(40,40)
%\dashline[30]{3}(40,40)(64,56)
%\dashline[30]{3}(40,0)(40,40)
\end{picture}\\*[6mm]

twisted 
\end{minipage}
\begin{minipage}[t]{40mm}
\centering
\begin{picture}(84,76)(-40,0)
\put(0,0){\circle*{4}}
\put(0,0){\line(3,2){24}}
\put(24,16){\circle*{4}}
\put(24,16){\line(0,1){40}}
\put(24,56){\circle*{4}}
\put(24,56){\line(1,0){40}}
\put(64,56){\circle*{4}}
\put(64,56){\line(0,-1){40}}
\put(64,16){\circle*{4}}
\put(64,16){\line(-3,-2){24}}
\put(40,0){\circle*{4}}
\put(40,0){\line(-1,0){40}}
% hidden lines 
\multiput(24,16)(4,0){11}{\circle*{0.1}}
%\dashline[30]{3}(24,16)(64,16)
\end{picture}\\*[6mm]

\ \ \ \ \ \ \ \ \ \ \ \ \ \ \ \  bent rectangle
\end{minipage}
\caption{Loops considered in the $O(a^2)$ on-shell tree-level Symanzik 
improvement of the 4d SU(N) lattice gauge theory: plaquette, 
$2\times 1$ rectangle, 
twisted perimeter-six loop and bent rectangle.}
\label{fig:loops}
\end{figure}
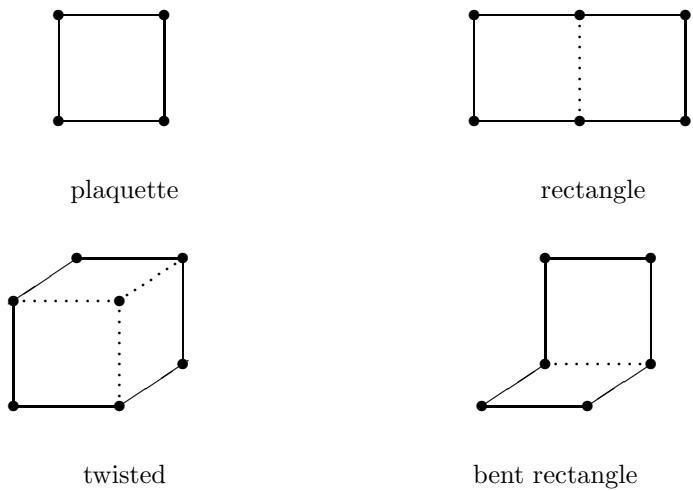

Following the notation of the Appendix of Ref.~\cite{FP97}, the lattice 
action for SU(N) can be written in a general form 
\beq
{\cal S}_L(U) = \be \: {\cal A}_L(U) \;\; , \;\;\;\;\; 
{\cal A}_L(U) = \f{1}{N}\sum_C c(C)[N-{\rm Re\:Tr}(U_C)] \;\; , \;\;\;\;\;
\be\equiv 2N/g^2 \;\;, 
\label{eq:act_sun}
\eeq
where $C$ denotes any closed path, $U_C$ stands for the product of the link
variables $U_\mu(n)\in$ SU(N) along the path $C$ and $c(C)$ is the coupling 
associated to the loop $C$ \footnote{Since we are interested at the lowest 
order in $a^2$, it is not necessary to consider higher powers of the trace in
Eq.~(\ref{eq:act_sun}).}. The lattice action can be re-expressed in terms 
of a symmetrized action density 
\beq
{\cal A}_L=\sum_n {\cal A}_L(n) \;\; ; \;\;\;\;\;
{\cal A}_L(n) = \f{1}{N} \sum_{C\ni n} c(C) \f{[N-{\rm Re\:Tr}(U_C)]}{{\rm
  perimeter}(C)} \;\;\;.  
\label{eq:dens_sun}
\eeq
Now, in order to expand ${\cal A}_L(n)$ in a power series of $a^2$, it is 
necessary to state the rule according to which the continuum gauge fields
$A_\mu^a(x)$, $a=1,\ldots,N^2-1$, are approximated by the link variables 
$U_\mu(n)$. This interpolation rule is somewhat arbitrary, although it is 
important that gauge covariance holds exactly, i.e. not only in the 
continuum limit $a\rightarrow 0$. A convenient choice is then
\beq
U_\mu(n)={\rm P}\exp \int_0^a A_\mu(na+s\hat\mu) \: {\rm d}s \;\;\; ,
\label{eq:rule_sun}
\eeq
where $A_\mu(x)=i\sum_a A^a_\mu(x) \lambda^a/2$ are anti-Hermitian gauge
fields (the $\lambda^a$ matrices, $a=1,\ldots,N^2-1$, are the generators of 
SU(N) in the fundamental representation). It can be easily proven, indeed, 
that the local gauge transformation of the link variables $U^\prime_\mu(n) 
\equiv g(na) U_\mu(n) g^\dagger (na+\hat\mu a)$ 
corresponds to the exact gauge transformation of the continuum fields 
$A_\mu^a(x)$. Using~(\ref{eq:rule_sun}), it is possible 
to expand ${\cal A}_L(n)$ in the form ${\cal A}_L(n)=
\sum^\infty_{k=0} a^{4+2k} {\cal O}^{(4+2k)}(na)$, where ${\cal O}^{(4+2k)}
(x)$ is a combination of gauge-invariant operators of the continuum with 
na\"{\i}ve dimension $4+2k$. Up to order $a^6$ this expansion reads
\bea
{\cal A}_L(n) &=& a^4 {\cal O}_0(na) \; \sum_C r_0(C) c(C) + a^6 
\left[ {\cal O}_1(na) \; \sum_C r_1(C)c(C) \right. \nonumber \\
  &+& \left. {\cal O}_2(na) \; \sum_C r_2(C)c(C) + {\cal O}_3(na) \;
  \sum_C r_3(C)c(C) \right] + O(a^8) \;\;\;, 
\label{eq:exp_sun} 
\eea
with 
\bea
{\cal O}_0(x) & = & -\f{1}{2} \sum_{\mu,\nu} {\rm
  Tr}(F^2_{\mu\nu}(x))\;\;\;, \\ 
{\cal O}_1(x) & = & \f{1}{12} \sum_{\mu,\nu} {\rm Tr}({\cal D}_\mu
F_{\mu\nu}(x))^2\;\;\;, \\  
{\cal O}_2(x) & = & \f{1}{12} \sum_{\mu,\nu,\lambda} {\rm Tr}({\cal D}_\mu
F_{\nu\lambda}(x))^2\;\;\;, \\
{\cal O}_3(x) & = & \f{1}{12}\sum_{\mu,\nu,\lambda} {\rm Tr}({\cal
  D}_\mu F_{\mu\lambda}(x){\cal D}_\nu F_{\nu\lambda}(x))\;\;\;, 
\eea
being ${\cal D}_\mu F_{\nu\lambda} \equiv \de_\mu F_{\nu\lambda}
+[A_\mu,F_{\nu\lambda}]$. The coefficients $r_0$, $r_1$, 
$r_2$ and $r_3$ are given in Table~2 of Ref.~\cite{FP97} for all the loops 
which live on a hypercube $2^4$. The normalization condition $\sum_C r_0(C) 
c(C)=1$ has to be satisfied in order to ensure the correct continuum limit. 
For the traditional choice of loops with perimeter not larger than six, 
i.e. the plaquette (pl), the $2\times 1$ rectangle (rt), the bent rectangle 
(br) and the twisted loop (tw) (see Fig.~\ref{fig:loops}), the 
expansion for ${\cal A}_L$ is
\bea
{\cal A}_L & = &  \int d^4 x\;\left\{ 
(c_{\rm pl}+8c_{\rm rt}+16c_{\rm br}+8c_{\rm tw})\;{\cal O}_0(x) 
\right.\nonumber \\
&+& a^2 \left[ (c_{\rm pl}+20c_{\rm rt}+4c_{\rm br}-4c_{\rm tw}) \; 
{\cal O}_1(x) \right. \label{eq:exp_sun_1} \\
&+& \left. \left. 4c_{\rm tw} \; {\cal O}_2(x) + (12c_{\rm br}+
4c_{\rm tw}) \; 
{\cal O}_3(x) \right] + O(a^4) \right\} \;\;\;, \nonumber
\eea
from which one reads immediately the normalization condition
$c_{\rm pl}+8c_{\rm rt}+16c_{\rm br}+8c_{\rm tw}=1$.

L\"uscher and Weisz observed in Ref.~\cite{LW85b} that the $O(a^2)$
on-shell tree-level Symanzik conditions determined through the calculation 
of on-shell scattering amplitudes~\cite{LW85a},
\beq
c_{\rm tw}=0\;\;,\;\;\;\;\; c_{\rm pl}+20c_{\rm rt}+4c_{\rm br}=0 \;\;,
\eeq
can be read off directly from Eq.~(\ref{eq:exp_sun_1}), by imposing that 
the coefficients of the operators ${\cal O}_1(x)$ and ${\cal O}_2(x)$ 
vanish. The fact that the term with ${\cal O}_3(x)$ plays no role, 
led them to observe that the {\em on-shell} improvement means the 
improvement of the action for {\em classical solutions} only. They
also declared in the footnote~(9) of Ref.~\cite{LW85b} that they
could not invert the reasoning because they did not know of an independent 
argument to this effect.

Another point which deserves a deeper understanding is the 
interpolation rule~(\ref{eq:rule_sun}): Why and to what extent is it 
arbitrary? Could one determine the $O(a^4)$ on-shell tree-level
Symanzik conditions by just making the lattice action more general and 
expanding in $a^2$, with the same interpolation rule, to the next order?
 
In the next Sections, it will be argued that FP actions
can provide a key to clarify these issues.

\section{Perfect actions}
\label{sec:perf}

In the following, the notation will be referred to the case 
of an unconstrained scalar field theory, in order to make the discussion 
more compact. The special cases of the 2d O(3) $\sigma$-model and of the 4d
SU(N) gauge theories will be treated shortly later on.

A RG transformation with scale factor 2 can be defined in the following way
\beq
\ex^{-\be^\prime{\cal A}_{L^\prime}(\Phi^\prime)} = \int D \Phi \:
\ex^{-\be \left[ {\cal A}_L(\Phi)+ \kappa \sum_n {\cal T}
(\Phi^\prime(n), f_\Phi(n))\right]}\;\;\;,
\label{eq:RG_2}
\eeq
where $\Phi$ is the lattice field living on the original lattice
with spacing $a $, $\Phi^\prime$ is the blocked lattice field living on 
a lattice with spacing $a^\prime = 2a $, ${\cal T}(\Phi^\prime(n),f_\Phi(n))$
is the blocking kernel, positive definite and normalized in order to
keep the partition function invariant under the transformation, 
$\kappa$ is an arbitrary parameter. The blocking kernel is chosen to be zero
when $\Phi^\prime(n)=f_\Phi(n)$: this relation, when applied to all the sites
$n$ of the blocked lattice, defines a blocking step. The functional 
$f_\Phi(n)$ is an averaging of the original field $\Phi$ in the 
surroundings of the site $n$ of the blocked lattice. Examples of RG
transformation with scale factor 2 are those defined by 
\beq
{\cal T}(\Phi^\prime(n),f_\Phi(n))=\left( \Phi^\prime(n) - f_\Phi(n) 
\right)^2 \;\;, 
\label{eq:ex1_1}
\eeq
or, in the $\kappa\rightarrow\infty$ limit, by
\beq
\ex^{ -\kappa {\cal T}(\Phi^\prime(n),f_\Phi(n))} \rightarrow
\delta\left( \Phi^\prime(n) - f_\Phi(n)\right )\;\; , 
\label{eq:ex1_2}
\eeq
with
\beq
f_\Phi(n)=\f{2^{\f{d-2}{2}}}{2^d}
\sum_{\lambda_1=\pm1/2}\cdots \sum_{\lambda_d=\pm1/2} 
\Phi(2n +\lambda_1 \hat 1 + \cdots + \lambda_d \hat d)\;\; .
\eeq
In the factor $2^{(d-2)/2}/2^d$, the denominator is the normalization of the 
average; the numerator appears because a scalar 
field in a $d$ space-time has na\"{\i}ve mass dimension equal to $(d-2)/2$,
and we are using instead fields in lattice units.  
In the cases in which the blocking step can be iterated analytically, it is
more convenient to perform one single RG transformation with scale factor 
equal to infinity between the continuum and a given lattice, namely
\beq
\ex^{-\be^\prime{\cal A}_L(\Phi)} = \int D \ph \:\ex^{-\be 
\left[ {\cal A}_{\rm cont}(\ph)
+ \kappa \sum_n {\cal T}(\Phi(n),f_\ph(n))\right]}\;\;\;,
\label{eq:RG}
\eeq
where now ${\cal A}_{\rm cont}(\ph)$ is the continuum action, 
functional of the 
continuum field $\ph$, and the field $\Phi$ is taken in physical
units\footnote{The continuum theory implies some 
regularization: the coupling $\be$ in Eq.~(\ref{eq:RG}) is the bare coupling 
at the regularization scale.}. The RG transformations with scale 
factor 2 defined by Eqs.~(\ref{eq:ex1_1}) 
and~(\ref{eq:ex1_2}) correspond to the following RG transformations with 
scale factor equal to infinity
\beq
{\cal T}(\Phi(n),f_\ph(n))=\left( \Phi(n) - f_\ph(n) \right)^2 
\;\;, 
\label{eq:ex1}
\eeq
and, in the $\kappa \rightarrow \infty$ limit, 
\beq
\ex^{-\kappa {\cal T}(\Phi(n),f_\ph(n))} \rightarrow
\delta\left( \Phi(n) - f_\ph(n)\right )\;\; , 
\label{eq:ex2}
\eeq
with
\beq
f_\ph(n)= \int_{-1/2}^{1/2}\cdots \int_{-1/2}^{1/2} d^d t \; \ph(na+ta) \;\;.
\eeq
In the following, we will restrict for simplicity to RG transformations with 
scale factor equal to infinity. The results can be easily extended to
all RG transformations with finite scale factor, by applying some trivial
iteration arguments. 

An important point to observe is that the integral 
in~(\ref{eq:RG}) is not definite for any RG transformation.
To have an intuition of this fact, let us 
consider the massless free scalar theory with the RG transformation 
defined in Eq.~(\ref{eq:ex2}): in Fourier transform, $\Phi(n) - f_\ph(n)$ 
can be written as
\beq
\Phi(p) - \sum_{l\in Z^d} \ph(p+2\pi l)\Pi(p+2\pi l)\;\;, \;\;\;\;\; 
\Pi(p)=\f{2\sin(p_\mu a/2)}{p_\mu}\;\;, 
\eeq
with $p$ in the first Brillouin zone $[-\pi/a,\pi/a]^d$. The 
integral~(\ref{eq:RG}) is quadratic and can be performed 
analytically giving
\beq
{\cal A}_L(\Phi)=\f{1}{2}\sum_{n,r} \Phi(n)\Phi(n+r)\rho(r)\;\;, 
\;\;\;\;\; \be^\prime=\be\;\;,
\eeq
with 
\beq
\f{1}{\rho(p)}=\sum_{l\in Z^d} \f{\Pi(p+2\pi l)^2}{(p+2\pi l)^2} + \f{1}
{2\kappa}\;\;.
\eeq
If the RG transformation were the ``decimation'', i.e. $f_\ph(n)=\ph(na)$, 
then everything would go the same way with the replacement $\Pi(p)
\rightarrow 1$, except that the summation involved in $1/\rho(p)$ would be 
no more convergent as soon as $d>1$. \newline

In all the cases of ``convergent'' RG transformations, the long
distance (in lattice units) properties of the lattice action 
${\cal A}_L(\Phi)$
defined by Eq.~(\ref{eq:RG}) are the same of the continuum action, without 
any cut-off dependence. The lattice action ${\cal A}_L(\Phi)$ is 
therefore a perfect action. 

For asymptotically free theories, since $\be^\prime=\be-O(1)$ in the 
$\be\rightarrow\infty$ limit, Eq.~(\ref{eq:RG}) can be solved 
by the saddle point approximation, giving
\beq
{\cal A}^{\rm FP}_L(\Phi)=\min_{\{\ph\}} \left[ {\cal A}_{\rm cont}(\ph) 
+ \kappa \sum_n {\cal T} (\Phi(n),f_\ph(n))\right]\;\;\;.
\label{eq:FP}
\eeq
This equation defines the FP action of any lattice configuration $\{\Phi\}$.
It is perfect only in the classical limit, i.e. at the tree-level.

Let us concentrate now on the classical solutions of a perfect lattice 
action.
A classical solution $\{\Phi_{\rm cl}\}$ is defined by the set of equations 
$\delta{\cal A}_L(\Phi)/\delta\Phi(m)=0$ or, equivalently, by 
\bea
& \f{\delta}{\delta\Phi(m)} \ex^{-\be^\prime {\cal A}_L(\Phi)} = 
\hspace{5cm} & \label{eq:eom} \\
& -\be\kappa \int D \ph \; \left[ \f{\delta}{\delta\Phi(m)}
{\cal T}(\Phi(m),f_\ph(m))\right] 
\ex^{-\be \left[ {\cal A}_{\rm cont}(\ph) + \kappa \sum_n 
{\cal T}(\Phi(n),f_\ph(n))) \right]}\;=0 \;\;. & \nonumber 
\eea
In the classical limit $\be\rightarrow\infty$, i.e. in the case of 
the FP action, the above equation can be easily solved. Indeed, for any 
fixed $\{\Phi\}$ there is a continuum configuration $\{\bar{\ph}\}$ such that
the integral in~(\ref{eq:eom}) can be approximated by the value of the 
integrand at $\{\ph\}= \{\bar{\ph}\}$. The configuration 
$\{\bar{\ph}\}$ is the minimum of the quantity in square brackets
at the exponent. In this limit, Eq.~(\ref{eq:eom}) reduces to 
$\f{\delta}{\delta\Phi(m)} {\cal T}(\Phi(m),f_{\bar{\ph}}(m))=0$. Being
${\cal T}(\Phi(m),f_\ph(m))$ positive definite with a local minimum 
at $\Phi(m)=f_\ph(m)$ where it is equal to zero, the solution of 
Eq.~(\ref{eq:eom}) is   
$\bar{\Phi}(m)=f_{\bar{\ph}}(m)$. Since 
${\cal T}(\bar{\Phi},f_{\bar{\ph}})=0$, $\{\bar{\ph}\}$
is a minimum of ${\cal A}_{\rm cont}(\ph)$ and 
is therefore a classical solution 
$\{\ph_{\rm cl}\}$ of the continuum equations of motion.
Summarizing, for any classical solution $\{\ph_{\rm cl}\}$ of the continuum 
theory there is a corresponding classical solution $\{\Phi_{\rm cl}\}$ 
of the FP action, related to the former by a blocking transformation.
It is easy to convince ourselves that ${\cal A}^{\rm FP}_L(\Phi_{\rm cl})=
{\cal A}_{\rm cont}(\ph_{\rm cl})$ also holds\footnote{For instanton classical
solutions in 2d models and in 4d SU(N), this is true up to a critical size
of the order of the lattice spacing (see, for instance, 
Ref.~\cite{BBHN96}).}. 
These results concerning FP classical 
solutions were obtained in~\cite{HN97,DHHN95a} in a 
slightly different notation.

\section{FP actions and on-shell tree-level Symanzik improvement}
\label{sec:fp_sym}

In the previous Section, it was shown that for any averaging functional 
$f_\ph(n)$ which defines a ``convergent'' RG transformation there is 
a related perfect lattice action, through Eq.~(\ref{eq:RG}). It was also
shown that the lattice solutions of the equations of motion 
in the classical limit can be put in correspondence with
the continuum solutions through $\{\Phi_{\rm cl}\}=\{f_{\ph_{\rm cl}}\}$, 
and that ${\cal A}^{\rm FP}_L(\Phi_{\rm cl})=
{\cal A}^{\rm FP}_L(f_{\ph_{\rm cl}})=
{\cal A}_{\rm cont}(\ph_{\rm cl})$. The latter chain of equations 
says that if we calculate
a FP lattice action on a classical solution $\{\Phi_{\rm cl}\}= 
\{f_{\ph_{\rm cl}}\}$ and
expand each $f_{\ph_{\rm cl}}(n)$ in power series of $a$, we must  
obtain that the leading term of this expansion reproduces exactly the 
continuum action 
calculated on $\{\ph_{\rm cl}\}$ and that {\em all} the irrelevant terms 
in $a^2$ vanish, at least by virtue of the continuum equations of motion.

Now, let us assume that we have built the FP action 
${\cal A}^{\rm FP}_L(\Phi)$ 
related to a certain averaging functional $f_\ph(n)$ and we calculate it
on the lattice field $\Phi(n)=f^\prime_{\ph_{\rm cl}}(n)$, where 
$f^\prime_\ph(n)$ is a different averaging functional, which is even allowed 
to define a non-convergent RG transformation. Since
\beq
{\cal A}^{\rm FP}_L(f^\prime_{\ph_{\rm cl}})={\cal A}_{\rm cont}
(\ph_{\rm cl})+
\sum_n \left. \f{1}{2}\f{\delta^2 {\cal A}^{\rm FP}_L(\Phi)}{\delta \Phi(n)^2}
\right|_{\{\Phi\} =\{f_{\ph_{\rm cl}}\}}
(f^\prime_{\ph_{\rm cl}}(n)- f_{\ph_{\rm cl}}(n))^2 + \cdots \;\;\;,
\label{eq:fp_act_exp}
\eeq
${\cal A}^{\rm FP}_L(f^\prime_{\ph_{\rm cl}})$ differs from the 
continuum action by terms
which are quadratic in the differences $f^\prime_{\ph_{\rm cl}}(n)
-f_{\ph_{\rm cl}}(n)$. By the 
very nature of average functional, both $f^\prime_\ph(n)$ and $f_\ph(n)$ can 
be expanded as $\ph(na)+O(a)$, so the sum of the differences 
$f^\prime_\ph(n)-f_\ph(n)$ represents a lattice correction 
to the continuum action of a classical solution. This correction will be 
$O(a^{2k})$, if the averaging functionals $f^\prime_\ph(n)$ and $f_\ph(n)$ 
differ by terms $O(a^k)$. 
This obvious result can be interpreted in the following sense: if one 
misses the correct rule which puts into correspondence the classical 
solutions
of a FP action with the classical solutions of the continuum theory, cut-off
effects come up at a certain predictable order in $a^2$. In the cases where 
a FP lattice action of a theory is known, this result has no 
practical use, except as a check. However, when no FP lattice actions of a 
theory have been built, but it is known by some
arguments that a certain averaging functional $f_\ph$ exists which {\em could}
define a FP action, the above result ensures that there exists at least {\em 
one} lattice action which is on-shell tree-level Symanzik improved at
$O(a^{2k})$ for any ``wrong'' functional $f^\prime_\ph$ which differs from
$f_\ph$ at the order $a^{k+1}$. A heuristic way to find this 
lattice action is the following:
\newline
1) start with a general form for the lattice action ${\cal A}_L(\Phi,c)$
dictated by the symmetry; $c$ is a collective index for the couplings,
constrained only to guarantee the correct continuum limit; \newline
2) take an averaging function $f^\prime_\ph$ (i.e. what we called the 
``interpolation rule'' in the Sect.~\ref{sec:sym}) which differs at the order
$a^{k+1}$ from an arbitrary averaging function $f_\ph$ which defines 
a ``convergent'' RG transformation;\newline
3) make the replacement $\Phi(n)=f^\prime_{\ph_{\rm cl}}(n)$ and expand in 
powers of $a$ up to the order $a^{2k}$ included, so that 
\beq
{\cal A}_L(\Phi,c) = {\cal A}_{\rm cont}(\ph_{\rm cl}) + \int d^dx \; 
\left[ \sum_{n=1}^{k} \sum_{m=1}^{M(n)} 
a^{2n} v^{(n)}_m(c) {\cal O}^{(n)}_m(\ph(x)) + O(a^{2k+2})\right ]\;\;, 
\eeq
where ${\cal O}^{(n)}_m(\ph)$, $m=1,\ldots,M(n)$, are continuum operators 
with na\"{\i}ve dimension $d+2n$; \newline
4) use the continuum equations of motion to reduce the number of the 
independent operators at any order $n$; \newline
5) impose $v^{(n)}_m(c)=0$, for $n=1,\ldots,k$ and $m=1,\ldots,M(n)$: these 
equations, together with the condition for the continuum limit, form 
the set of the Symanzik conditions for the on-shell tree-level improvement 
to $O(a^{2k})$.

The only non-trivial task in the above recipe is to find an averaging
functional $f_\ph$ which is known from independent arguments to define
a convergent RG transformation\footnote{In many cases, it is sufficient to 
check that there is convergence for the free theory.}.

As an illustration of the above procedure, let us consider the 2d O(3) 
$\sigma$-model, for which the FP action has been built in 
Ref.~\cite{HN94} by iterating the RG transformation with scale factor 
2 defined by the average functional
\beq
f_{\vS}(n)= \f{\sum_{\lambda_1=\pm 1/2}\sum_{\lambda_2=\pm 1/2} 
\vS_{2n+\lambda_1\hat 1+\lambda_2\hat 2}}
{\left| \sum_{\lambda_1=\pm 1/2}\sum_{\lambda_2=\pm 1/2} 
\vS_{2n+\lambda_1\hat 1+\lambda_2\hat 2} \right|}
\;\;.
\eeq

The lattice action can be written in a general form\footnote{The notation 
is that of Ref.~\cite{HN97}.} 
\bea
{\cal S}_L &=& \be \: {\cal A}_L\;\;, \;\;\;\;\; \be\equiv1/g  \nonumber \\
{\cal A}_L &=& -\f{1}{2}\sum_{n_1,n_2} \rho(n_1-n_2)(1-\vS_{n_1}
\cdot\vS_{n_2}) \\
&+& \sum_{n_1,n_2,n_3,n_4} c(n_1,n_2,n_3,n_4)(1-\vS_{n_1}\cdot\vS_{n_2})
(1-\vS_{n_3}\cdot\vS_{n_4}) + \ldots \;\;\;, \nonumber  
\label{eq:act_o3}
\eea 
with the constraint $\vS_{n}\cdot\vS_{n} =1$. 
If we are interested in the on-shell Symanzik improvement at $O(a^2)$, it is
sufficient to choose\footnote{This averaging would define the non-convergent 
RG transformation of decimation.}
\beq
f^\prime_{\vS}(n) = \vS_n \;\; ,
\eeq
since $f_\vS$ and $f^\prime_\vS$ differ at the order $a^2$. 
The analog for scale factor equal to infinity would be $f^\prime_{\vS}(n)=
\vS(na)$, where $\vS(x)$ at the r.h.s. is the continuum field. Replacing 
now $\vS_n$ with $\vS(na)$ in ${\cal A}_L$ and expanding in powers of $a^2$,
one finds~\cite{HN97}
\bea
{\cal A}_L &=& \int d^2x \:\left\{ \f{1}{2} \de_\mu\vS\cdot\de_\mu\vS \right.
\nonumber \\
&+& a^2\left[\f{R_1}{16}\,(\de^2\vS\cdot\de^2\vS)+\f{R_2}{48}\sum_\mu 
(\vS\cdot\de_\mu^4\vS) \right. + \f{C_1}{4}(\vS\cdot\de^2\vS)^2 
\label{eq:exp_o3} \\ 
&+& \left.\left.\f{C_2}{2} \sum_{\mu,\nu}
(\de_\mu\vS\cdot\de_\nu\vS)^2 +\f{C_3}{4}\sum_\mu(\de_\mu\vS\cdot\de_\mu\vS)^2
\right] + O(a^4)\right\}\;\;, 
\nonumber
\eea
having defined
\beq
\sum_n \rho(n)n_\mu n_\nu n_\alpha n_\be = R_1 (\delta_{\mu\nu}\delta_{\alpha
\be} + \delta_{\mu\alpha}\delta_{\nu\be} + \delta_{\mu\be}\delta_{\nu\alpha})
+ R_2\;\delta_{\mu\nu\alpha\be}
\eeq
and
\bea
\f{1}{V}& \!\!\!\!\!\!\!\!\!\!\!\!\!\!
\sum_{n_1,n_2,n_3,n_4} c(n_1,n_2,n_3,n_4) \Delta_\mu \Delta_\nu
\Delta^\prime_\alpha \Delta^\prime_\be = 
\hspace{2cm} & \nonumber \\
& \hspace{2cm} C_1\:\delta_{\mu\nu}\delta_{\alpha\be}
+ C_2\:(\delta_{\mu\alpha}\delta_{\nu\be} + 
\delta_{\mu\be}\delta_{\nu\alpha}) + C_3 \: \delta_{\mu\nu\alpha\be}\;\;, &
\eea
where $\Delta=n_1-n_2$, $\Delta^\prime=n_3-n_4$ and 
$\delta_{\mu\nu\alpha\be}$
is 1 when all its indices are equal, otherwise it is zero. Using the 
equations of motion $\de^2\vS=\vS(\vS\cdot \de^2\vS)$, the $O(a^2)$ on-shell 
tree-level Symanzik conditions can be read off from Eq.~(\ref{eq:exp_o3})
\beq
R_2=0\;\;\;, \;\;\;\;\; C_1+\f{1}{4} R_1=0\;\;\;,\;\;\;\;\;C_2=0\;\;\;,
\;\;\;\;\;C_3=0\;\;\;.
\eeq

To improve the action at $O(a^4)$, $f^\prime_\vS$ should be chosen to 
differ from $f_\vS$ at $O(a^3)$, and so on. Of course, the new 
$f^\prime_\vS$ should be chosen simple 
enough that it can be analytically iterated an infinite number of times.
Alternatively, one could work with scale factor equal to infinity from the 
beginning and take for instance
\beq
f_{\vS}(n)= \f{\int_{-1/2}^{1/2}  \int_{-1/2}^{1/2} d^2t \: 
\vS(na+ta)}{\left|\int_{-1/2}^{1/2}  \int_{-1/2}^{1/2} d^2t \: 
\vS(na+ta)\right|} \;\;,
\eeq
and choose $f^\prime_\vS$ consequently. Of course, the averaging functional
$f_{\vS}$ would define a different FP action, which however needs not to be 
determined in the context of the Symanzik improvement.

In view of the above considerations, the procedure followed in 
Sect.~\ref{sec:sym} for the $O(a^2)$ improvement in 4d SU(N) gauge theories
should be fully under control. In that case, the 
interpolation rule adopted in Refs.~\cite{GGSV94,FP97} corresponds to
the iteration of the scale factor 2 blocking transformation defined by 
\beq
f^\prime_{U_\mu}(n) \; = \; U_\mu(2n) \: U_\mu(2n+\hat\mu) \;\;.
\eeq
In four dimensions this averaging functional does not define a 
``convergent'' RG transformation. We know, however, that there exist 
averaging functionals which allow to build FP actions for
SU(N)~\cite{DHHN95a,BN96}. The functional $f^\prime_{U_\mu}$ differs
from those averaging functionals at the $O(a^2)$, at the level of the gauge
fields $A_\mu(n)$. So, the same arguments
apply as in the case of the O(3) $\sigma$-model\footnote{Going to the next 
order in $a^2$ in gauge theories with this method, 
however, involves serious technical problems owing to the limitations posed 
by gauge invariance.}.

A point which should be remarked is that, both in the case of the 2d O(3)
$\sigma$ model and of 4d SU(N) gauge theories, the $O(a^2)$ on-shell 
tree-level Symanzik improvement results in an {\em infinity} of possible
lattice actions. The argument described in this Section ensures that 
for any averaging functionals $f_\ph$ and $f^\prime_\ph$ there is 
only {\em one} on-shell tree-level Symanzik improved action to a given 
order in $a^2$. This could either reflect the fact that for any 
fixed $f^\prime_\ph$ 
there can be an infinite number of ``good'' functionals $f_\ph$ defining
FP lattice actions, or that FP actions are a sub-class of all
the lattice actions which satisfy ${\cal A}_L(f_{\ph_{\rm cl}})=
{\cal A}_{\rm cont}(\ph_{\rm cl})$ and that this last condition
is sufficient to ensure the on-shell tree-level Symanzik improvement
to all orders in $a^2$. Although something remain to be understood, FP
actions provide anyway an interesting approach to the on-shell tree-level 
Symanzik improvement.

\section{Acknowledgments} 

This work stemmed from a discussion with Federico Farchioni at the time of 
our work of Ref.~\cite{FP97}. I am indebted to him also for critically
reading the manuscript and for the subsequent stimulating e-mail 
correspondence. I acknowledge also an interesting conversation with Matteo 
Beccaria.

\end{document}